\documentclass[graybox]{svmult}

\usepackage{mathptmx}       % selects Times Roman as basic font   maha se za po-krasivo ...
\usepackage{helvet}         % selects Helvetica as sans-serif font
\usepackage{courier}        % selects Courier as typewriter font
\usepackage{type1cm}        % activate if the above 3 fonts are
\usepackage{makeidx}         % allows index generation
\usepackage{graphicx}        % standard LaTeX graphics tool
\usepackage{multicol}        % used for the two-column index
\usepackage[bottom]{footmisc}% places footnotes at page bottom

\usepackage{cite}
\begin{document}

\title*{Group analysis of a class of nonlinear Kolmogorov equations}
% Use \titlerunning{Short Title} for an abbreviated version of
% your contribution title if the original one is too long
\author{Olena~Vaneeva, Yuri Karadzhov and Christodoulos Sophocleous}
% Use \authorrunning{Short Title} for an abbreviated version of
% your contribution title if the original one is too long
\institute{Olena~Vaneeva \at Institute of Mathematics of the National Academy of Sciences of Ukraine, 3 Tereshchenkivska Str., Kyiv-4, 01601 Ukraine \email{vaneeva@imath.kiev.ua}
\and Yuri Karadzhov \at Institute of Mathematics of the National Academy of Sciences of Ukraine, 3 Tereshchenkivska Str., Kyiv-4, 01601 Ukraine \email{yuri.karadzhov@gmail.com}
\and Christodoulos Sophocleous \at Department of Mathematics and Statistics, University of Cyprus, Nicosia CY 1678, Cyprus \email{christod@ucy.ac.cym}
}
\maketitle

\abstract{A class of $(1{+}2)$-dimensional diffusion-convection equations
(nonlinear Kolmogorov equations) with time-dependent coefficients
is studied with Lie symmetry point of view. The complete group classification
is achieved using a~gauging of arbitrary elements
(i.e. via reducing the number of variable coefficients) with the application of equivalence transformations.
Two possible gaugings are discussed in order to show how equivalence groups
serve in making the optimal  choice.}

\section{Introduction}
Second-order partial differential equations of the form
\begin{eqnarray}\label{eq_2dDiff0}
u_t= D u_{yy}+\nu \left[K(u)\right]_x,
\end{eqnarray}
where $D$ and $\nu$ are nonzero constants, and $K$ is a smooth nonlinear function of the dependent variable $u$, appear in various applications. In particular, they describe diffusion-convection processes~\cite{Escobedo}, model an interaction of particles with two kinds of particles on a lattice~\cite{Alexander}, arise in mathematical finance, when studying agents' decisions
under risk~\cite{Citti,Pascucci}. Equations~(\ref{eq_2dDiff0}) are called in the literature diffusion-advection equations, nonlinear ultraparabolic equations and nonlinear Kolmogorov equations. They were studied from various points of view. An important study of partial differential equations and especially nonlinear ones is finding Lie groups of point transformations that leave an equation under study invariant.
Such symmetry transformations allow one to apply powerful, and
what is most important, algorithmic methods
for finding exact solutions  of a given nonlinear equation. Moreover, Lie symmetries can serve as a selection criterion of physically important models among possible ones~\cite{FN}.
Lie symmetries of  equations~(\ref{eq_2dDiff0}) and the corresponding group invariant solutions were classified by Demetriou et al~\cite{Demetriou}. There are also studies on Lie symmetries of linear Kolmogorov equations~\cite{Kov1,Kov2} and of constant coefficient nonlinear Kolmogorov equations of the form $u_t- u_{yy}-uu_x=f(u)$~\cite{Ser}.

An attempt of group classification of a class of nonlinear Kolmogorov equations more general than~(\ref{eq_2dDiff0}),
namely, such equations with time dependent coefficients,
\begin{equation}\label{eq_2dDiff}
u_t=f (t)u_{yy}-g (t)\!\left[K(u)\right]_x, \quad f g K_{uu}\neq0,
\end{equation}
was recently made~\cite{Kumar2014}. Here $f$  and $g$ are smooth nonvanishing functions of the variable $t,$ and $K$ is a smooth nonlinear function of $u.$
Nevertheless the complete classification of Lie symmetries of class~(\ref{eq_2dDiff}) was not achieved  in~\cite{Kumar2014}, in particular, the case $K=u\ln u$ was missed and dimensions of maximal Lie symmetry algebras as well as some of their basis elements for the other cases of extensions were presented incorrectly. The case $K=u^2$ that is
important for applications was not studied with Lie symmetry point of view at all.

In this paper we perform the complete group classification of equations~(\ref{eq_2dDiff}). As class~(\ref{eq_2dDiff})  is parameterized by three arbitrary elements, $K(u)$, $f(t)$ and $g(t)$,
the group classification problem appears to be too complicated to be solved completely without modern approaches based on the usage of point equivalence transformations. One of such tools  is the gauging of arbitrary elements by equivalence transformations
(i.e., reducing of a class to a subclass with fewer number of arbitrary elements).
To use this technique, we firstly compute the equivalence group of
class~(\ref{eq_2dDiff}) in Section 2.
A gauging of arbitrary elements is performed in the same section. In Section~3 Lie symmetries of the simplified class are exhaustively classified.
In Section~4 we discuss how to choose an optimal gauging among possible ones. To illustrate that the chosen gauging is optimal, we also present results on group classification of class~(\ref{eq_2dDiff}) carried out for an alternative gauging.

\section{Equivalence transformations}

Equivalence transformations are nondegenerate point transformations, that preserve the differential structure of the  class under study, change only its arbitrary elements and form a group. There are several kinds of equivalence groups. The {\it usual equivalence group}, used   for solving group classification problems since the late 50's,  consists of the nondegenerate point transformations
of the independent and dependent variables and of the arbitrary elements of the class,
where transformations for independent and dependent variables do not involve arbitrary elements of the class~\cite{Ovsiannikov1982}. The notion of
 the {\it generalized equivalence group}, where transformations of variables of given DEs explicitly depend on arbitrary elements, appeared  in the middle 90's~\cite{Meleshko1994,Meleshko1996}.
The transformations from the {\it extended equivalence group}
 include nonlocalities with respect to arbitrary elements~\cite{mogran}.
The {\it  generalized extended equivalence
group} possesses  the properties of both generalized and extended equivalence groups.
The group classification problems become simpler for solving if one use the widest possible equivalence group. Advantages of the usage of the generalized extended equivalence group in comparison with the usual one were shown, in particular, in~\cite{VKS2015}. In some cases the usage of generalized extended equivalence groups is the only way to present the complete group classification, see, e.g.,~\cite{VPS2012}.

Equivalence transformations generate a subset of a set of admissible transformations
which can be interpreted as  triples, each of which consists
of two fixed equations from a class and a point transformation that links these two equations~\cite{popo2010a,popo2012}.
In this paper we restrict ourselves to the study of equivalence transformations.

To find the equivalence transformations we use the direct method~\cite{Kingston&Sophocleous1998}.
The details of calculations are skipped for brevity. As it is more convenient for the study of Lie symmetries to consider the equivalent form of the above class,
\begin{eqnarray}\label{eq_2dDiff2}
u_t=f(t)u_{yy}-g(t)k(u)u_x, \quad f g k_u\neq0,
\end{eqnarray}
we present transformations for both $K$ and $k=K_u$ in the theorems below.
\begin{theorem} The generalized extended equivalence group~$\hat G^{\sim}$ of class~(\ref{eq_2dDiff}) (resp.~(\ref{eq_2dDiff2})) is formed by the transformations
\begin{eqnarray*}
&\tilde t=T(t),\quad \tilde x=\delta_1 x+\delta_2\int\!  g (t)\, {\rm d}t+\delta_3,\quad \tilde y=\delta_4y+\delta_5,\quad
\tilde u=\delta_6 u+\delta_7, \\
&\displaystyle\tilde f(\tilde t)=\frac{{\delta_4}^2}{T_t}f(t),\quad\tilde g (\tilde t)=\frac{\varepsilon_1}{T_t}g (t),\\
 &\displaystyle\tilde K(\tilde u)=\frac{\delta_6}{\varepsilon_1}\left(\delta_1K(u)+\delta_2u+\varepsilon_2\right),\quad
\left(\mbox{resp.}\quad \tilde k(\tilde u)=\frac1{\varepsilon_1}(\delta_1k(u)+\delta_2),\right)
\end{eqnarray*}
where  $\delta_i,$ $i=1,\dots,7,$ $\varepsilon_1$ and $\varepsilon_2$ are arbitrary constants with
$\delta_1\delta_4\delta_6\varepsilon_1\not=0$, $T(t)$ is an arbitrary smooth function with $T_t\neq0.$

The usual equivalence group of class~(\ref{eq_2dDiff}) (resp.~(\ref{eq_2dDiff2})) consists of the above transformations with $\delta_2=0.$
\end{theorem}

The group $\hat G^{\sim}$ contains a subgroup of gauge equivalence transformations, i.e. the transformations that change only arbitrary elements while the independent and dependent variables remain unchanged~\cite{popo2010a}. This  subgroup is formed by the transformations $\tilde t=t,$ $\tilde x=x$, $\tilde y=y,$ $\tilde u=u,$  $\tilde f=f,$ $\tilde g=\varepsilon_1 g$, $\tilde K=(K+\varepsilon_2)/\varepsilon_1$ (resp. $\tilde k= k/\varepsilon_1$). It is more convenient to consider  class~(\ref{eq_2dDiff2}) than class~(\ref{eq_2dDiff}) as in this case the dimension of the gauge equivalence subgroup reduces.

It appears that the subclass of equations~(\ref{eq_2dDiff}) with  $K$  quadratic  in $u$ (resp.~(\ref{eq_2dDiff2}) with $k$  linear in $u$)  admits a~wider equivalence group. Up to the $\hat G^{\sim}$-equivalence we can consider the case $K=u^2$ (resp. $k=u$).
\begin{theorem}The generalized extended equivalence group~$\hat G^{\sim}_1$ of the class
\begin{equation}\label{eq_2dDiff3}
u_t=f(t)u_{yy}-g(t)uu_{x}, \quad fg  \neq0,
\end{equation}
 comprises  the transformations
\begin{eqnarray*}&\tilde t=T(t),\quad \tilde x=X (t)x+\delta_3\!\int\!\! g(t)X (t)^2{\rm d}t+\delta_4,\quad \tilde y=\delta_1y+\delta_2,\\&\displaystyle
 \tilde u=\delta_5\!\left(\frac{u}{X (t)}- \delta_6x+\delta_3\right)\!,\quad \tilde f (\tilde t)=\frac{\delta_1^2}{\delta_5T_t}f (t),\quad \tilde g (\tilde t)=\frac{X (t)^2}{\delta_5T_t}g (t),
\end{eqnarray*}
where  $ X (t)=1/\left({\delta_6\int\! g(t)\, {\rm d}t+\delta_7}\right),$ $\delta_i,$ $i=1,\dots,7,$ are arbitrary constants with
$\delta_1\delta_5(\delta_6^2+\delta_7^2)\not=0$, and $T(t)$ is an arbitrary smooth function with $T_t\neq0.$

The usual equivalence group of class~(\ref{eq_2dDiff3}) consists of the above transformations with \mbox{$\delta_3=\delta_6=0.$}
\end{theorem}

As there is one arbitrary function, $T(t)$, in the transformations from the group~$\hat G^{\sim}$, we can set one of the arbitrary elements $f$ or $g$ of the initial  class equals to a~nonzero constant value.
We choose to perform the gauging $g=1$ by using the  transformation
\begin{equation}\label{tr}\textstyle \tilde t=\int\! g (t)\,{\rm d}t, \quad \tilde x=x,\quad\tilde u=u.\end{equation} Then, any equation from class~(\ref{eq_2dDiff}) (resp.~(\ref{eq_2dDiff2})) is mapped to an equation from its subclass that is singled out by the condition  $g =1$. The detailed discussion on optimal choice of gauging is presented in Section 4.
Without loss of generality, we can restrict ourselves to the study of  class~(\ref{eq_2dDiff}) with $g=1$ or, what is more convenient, its equivalent form
\begin{eqnarray}\label{eq_2dDiff4}
u_t=f (t)u_{yy}-k(u)u_x,\quad f k_u\neq0,
\end{eqnarray}
since all results on symmetries, conservation laws, classical solutions and other related objects can be found for equations~(\ref{eq_2dDiff2})
using the similar results derived for~(\ref{eq_2dDiff4}).

 The generalized extended equivalence groups of class~(\ref{eq_2dDiff4}) and its subclass with $k=u$ coincide with the usual equivalence groups of these classes.
\begin{theorem}The usual equivalence group~$G^{\sim}$ of class~(\ref{eq_2dDiff4}) consists of  the transformations
\begin{eqnarray*}
&\tilde t=\varepsilon_1t+\varepsilon_0,\quad \tilde x=\delta_1 x+\delta_2t+\delta_3,\quad \tilde y=\delta_4y+\delta_5,\quad
\tilde u=\delta_6 u+\delta_7, \\&\displaystyle
\tilde f (\tilde t)=\frac{{\delta_4}^2}{\varepsilon_1}f (t),\quad\tilde k(\tilde u)=\frac1{\varepsilon_1}(\delta_1k(u)+\delta_2),
\end{eqnarray*}
where  $\delta_i,$ $i=1,\dots,7$, $\varepsilon_1$
 and $\varepsilon_0$ are arbitrary constants with
$\delta_1\delta_4\delta_6\varepsilon_1\not=0$.
\end{theorem}
\begin{theorem}The usual equivalence group~$G^{\sim}_1$ of the class
\begin{equation}\label{eq_2dDiff5}
u_t=f(t)u_{yy}-uu_{x}, \quad f\neq0,
\end{equation}
is formed by  the transformations
\begin{eqnarray*}
&\displaystyle \tilde t=\frac{\alpha t+\beta}{\gamma t+\delta},\quad \tilde x=\frac{\kappa x+\mu t+\nu}{\gamma t+\delta},\quad \tilde y=\lambda y+\varepsilon,\\&\displaystyle
 \tilde u=\frac1{\Delta}\left(\kappa(\gamma t+\delta)u-\kappa\gamma x+\delta\mu-\gamma\nu\right),\quad \tilde f(\tilde t)=\frac{\lambda^2}{\Delta}(\gamma t+\delta)^2f(t),
\end{eqnarray*}
where $\alpha,$ $\beta,$ $\gamma,$ $\delta,$ $\kappa,$ $\mu,$ and $\nu$ are arbitrary constants defined up to a nonzero multiplier with
$\Delta=\alpha\delta-\beta\gamma\not=0$, $\kappa\neq0$; $\lambda$ and $\varepsilon$ are arbitrary constants, $\lambda\neq0$.
\end{theorem}
 Theorem~4 implies that any equation~(\ref{eq_2dDiff5}) with $f=a(t+b)^{-2}$, where $a\neq0$ and $b$ are constants, is  mapped by a point transformation to a constant-coefficient equation from the same class.

We also present  equivalence transformations for the subclass of class~(\ref{eq_2dDiff2}) singled out by the condition  $f=1$, which we will use for the comparison of the cases $f=1$ and $g=1$ in Section 4.
\begin{theorem}The generalized extended equivalence group~$\hat {G}^\sim_2$ of the class
\begin{equation}\label{eq_2dDiff7}
u_t=u_{yy}-g(t)k(u)u_{x}, \quad gk_u \neq0,
\end{equation} comprises  the transformations
\begin{eqnarray*}
&\tilde t=\delta_4^2t+\delta_0,\quad \tilde x=\delta_1 x+\delta_2\int\!  g(t)\,{\rm d}t+\delta_3,\quad \tilde y=\delta_4y+\delta_5,\quad
\tilde u=\delta_6 u+\delta_7, \\&\displaystyle
\tilde g(\tilde t)=\frac{\varepsilon_1}{\delta_4^2}g(t),\quad\tilde k(\tilde u)=\frac{1}{\varepsilon_1}\left(\delta_1k(u)+\delta_2\right),
\end{eqnarray*}
where  $\delta_i,$ $i=0,1,\dots,7,$ and $\varepsilon_1$  are arbitrary constants with
$\delta_1\delta_4\delta_6\varepsilon_1\not=0$.
\end{theorem}
\begin{theorem}The generalized extended equivalence group~$\hat G^{\sim}_3$ of the class
\begin{equation}\label{eq_2dDiff6}
u_t=u_{yy}-g(t)uu_{x}, \quad g \neq0,
\end{equation}
 consists of  the transformations
\begin{eqnarray*}&\displaystyle
\tilde t=\delta_1^2t+\delta_2,\quad \tilde x=\frac {x+\delta_4}{\gamma_1\int\!   g (t)\,{\rm d}t+\gamma_2}+\delta_5,\quad \tilde y=\delta_1y+\delta_3,  \\&
 \tilde u=\delta_6\!\left(\!\left(\gamma_1\!\int\!   g (t)\,{\rm d}t+\gamma_2\right) u-\gamma_1(x+\delta_4)\!\right),\displaystyle\quad\tilde  g (\tilde t)=\frac{ g (t)}{\delta_1^2\delta_6\!\left(\gamma_1\int\!   g (t){\rm d}t+\gamma_2\right)^2},
\end{eqnarray*}
where  $\delta_i,$ $i=1,\dots,6,$ $\gamma_1$ and $\gamma_2$ are arbitrary constants with
$\delta_1\delta_6(\gamma_1^2+\gamma_2^2)\not=0$.
\end{theorem}

%In the next section we demonstrate usage of the derived equivalence transformations  in the process of group classification.

\section{Lie symmetries}
The group classification problem for class~(\ref{eq_2dDiff2}) up to $\hat G^{\sim}$-equivalence reduces to
the similar problem for class~(\ref{eq_2dDiff4}) up to $G^{\sim}$-equivalence (resp. the group classification problem for class~(\ref{eq_2dDiff3}) up to $\hat G_1^{\sim}$-equivalence reduces to
such a problem for  class~(\ref{eq_2dDiff5}) up to $G_1^{\sim}$-equivalence).

To solve the group classification problem for class~(\ref{eq_2dDiff4}) we use
the classical approach based on integration of determining equations implied by the infinitesimal invariance criterion~\cite{Ovsiannikov1982}.
We
search for symmetry operators of the form $Q=\tau(t,x,y,u)\partial_t+\xi(t,x,y,u)\partial_x+\eta(t,x,y,u)\partial_y+\theta(t,x,y,u)\partial_u$  generating one-parameter Lie groups
of transformations that leave equations~(\ref{eq_2dDiff4}) invariant~\cite{Olver1986,Ovsiannikov1982}.
It is required that the action of the second prolongation $Q^{(2)}$ of the operator~$Q$ on~(\ref{eq_2dDiff4}) vanishes identically modulo equation~(\ref{eq_2dDiff4}),
\begin{equation}\label{c2}
Q^{(2)}\{u_t- f (t)u_{yy}+k(u)u_{x}\}|_{u_t=f (t)u_{yy}-k(u)u_x}=0.
\end{equation}

The infinitesimal invariance criterion~(\ref{c2}) implies the determining equations, simplest of which result in
\begin{eqnarray*}
\tau=\tau(t),\quad
\xi=\xi(t,x), \quad \eta=\eta^1(t)y+\eta^0(t),\quad
\theta=\varphi(t,x,y)u+\psi(t,x,y),
\end{eqnarray*}
where $\tau$, $\xi$, $\eta^1$, $\eta^0$, $\varphi$ and $\psi$ are arbitrary smooth functions of their variables.
Then the rest of the determining equations are
\begin{eqnarray}&
\tau f_t=(2\eta^1-\tau_t)f,\quad
2f\varphi_y=-\eta^1_t y-\eta^0_t,\label{deteq1}\\&
(\varphi u+\psi)  k_u+(\tau_t-\xi_x)k=\xi_t,\label{deteq2}\\&
(\varphi_x u+\psi_x)  k +(\varphi_t-f\varphi_{yy})u+\psi_t-f\psi_{yy}=0.\label{deteq3}
\end{eqnarray}

Firstly we integrate equations~(\ref{deteq2}) and~(\ref{deteq3}) for $k$ up to the $G^\sim$-equivalence taking into account that $k_u\neq0$.
 The method of furcate split~\cite{Nikitin&Popovych2001,Ivanova&Popovych&Sophocleous2006I} is further used.
 For any operator $Q\in A^{\rm max}$ equation~(\ref{deteq2})  gives   equations on $k$
of the general form
\begin{eqnarray}\label{eq_k}
(au+b)k_u+ck=d,
\end{eqnarray}
where $a,$ $b,$ $c,$ and $d$ are constants. The number $s$ of such independent equations is not greater than two, otherwise they form incompatible system for $k.$ If $s=0$, then~(\ref{eq_k}) is not an equation on $k$ but an identity, this corresponds to the case of arbitrary $k$.
If $s=1$, then the integration of~(\ref{eq_k}) up to the $G^\sim$-equivalence gives three different cases: (i) $k=u^n$, $n\neq0,1;$ (ii) $k=e^u;$ (iii) $k=\ln u.$
If $s=2,$ then the function $k$ is linear in $u$, $k=u\bmod G^\sim.$

The determining equation~(\ref{deteq3}) implies that there exist two essentially different cases of classification:
I. $k_{uu}\neq0$, and II. $k_{uu}=0$, i.e. $k=u\bmod G^\sim.$

Consider firstly the case of arbitrary function $k$.
In this case  equations~(\ref{deteq2}) and~(\ref{deteq3}) should be split with respect to $k$ and $k_u$.
The splitting results in the equations $\varphi=\psi=\xi_t=\tau_t-\xi_x=0.$ Therefore  $\tau=c_1t+c_2,$ $\xi=c_1x+c_3$.
As $\varphi=0$,  the second equation of~(\ref{deteq1}) implies $\eta^1_t=\eta^0_t=0,$ i.e. $\eta^1=c_4,$  and $\eta^0=c_5.$ Here $c_i,$ $i=1,\dots,5,$ are arbitrary constants.
Then the general form of the infinitesimal generator is
$
Q=(c_1t+c_2)\partial_t+(c_1x+c_3)\partial_x+(c_4y+c_5)\partial_y
$
and the first equation of~(\ref{deteq1}) takes the form
\begin{equation}\label{classeq1}
(c_1t+c_2)f_t=(2c_4- c_1)f.
\end{equation}
This is the classifying equation for $f.$
If $f$ is an arbitrary nonvanishing smooth function, then the latter equation should be split with respect to $f$ and its derivative, which results in $c_1=c_2=c_4=0.$
Therefore, the kernel $A^\cap$ of the maximal Lie invariance algebras of equations from class~(\ref{eq_2dDiff4})
is $A^\cap=\langle\partial_x,\,\partial_y\rangle$ (Case~1 of Table~1).
To perform the further classification we integrate equation~(\ref{classeq1}) up to the $G^{\sim}$-equivalence.
All $G^\sim$-inequivalent values of $f$ that provide  Lie symmetry extensions  for equations from  class~(\ref{eq_2dDiff4}) with arbitrary $k$ are exhausted by the following values:
 $f =t^\rho, \  \rho\neq0;$ $f =e^t;$ $f =1.$ The corresponding bases of maximal Lie invariance algebras are presented by Cases~2--4 of Table~1.

\begin{table}[t!]\centering
\caption{The group classification of class~(\ref{eq_2dDiff4}) up to the $G^\sim$-equivalence.}
\begin{tabular}{ccl}
\hline\noalign{\smallskip}
no.&$f (t)$&\hfil Basis of $A^{\max}$ \\
\noalign{\smallskip}\svhline\noalign{\smallskip}
\multicolumn{3}{c}{Arbitrary $k$}\\
\hline\noalign{\smallskip}
1&
$\forall$&$\partial_x,\quad\partial_y$\\{\smallskip}

2&
$t^\rho$&$\partial_x,\quad\partial_y,\quad2t\partial_t+2x\partial_x+(\rho+1)y\partial_y$\\{\smallskip}

3&
$e^t$&$\partial_x,\quad\partial_y,\quad2\partial_t+y\partial_y$\\{\smallskip}

4&
$1$&$\partial_x,\quad\partial_y,\quad\partial_t,\quad2t\partial_t+2x\partial_x+y\partial_y$\\
\hline\noalign{\smallskip}
\multicolumn{3}{c}{$k=u^{n}$, ${n}\neq0,1$}\\
\hline\noalign{\smallskip}
5&
$\forall$&$\partial_x,\quad\partial_y,\quad{n} x\partial_x+u\partial_u$\\{\smallskip}

6&
$t^\rho$&$\partial_x,\quad\partial_y,\quad{n} x\partial_x+u\partial_u,\quad2 t\partial_t+2x\partial_x+(\rho+1)y\partial_y$\\{\smallskip}

7&
$e^t$&$\partial_x,\quad\partial_y,\quad{n} x\partial_x+u\partial_u,\quad2\partial_t+y\partial_y$\\{\smallskip}

8&
1&$\partial_x,\quad\partial_y,\quad{n} x\partial_x+u\partial_u,\quad\partial_t,\quad2t\partial_t+2x\partial_x+y\partial_y$\\
\hline\noalign{\smallskip}
\multicolumn{3}{c}{$k=e^u$}\\
\hline\noalign{\smallskip}
9&
$\forall$&$\partial_x,\quad\partial_y,\quad  x\partial_x+\partial_u$\\{\smallskip}

10&
$t^\rho$&$\partial_x,\quad\partial_y,\quad  x\partial_x+\partial_u,\quad2 t\partial_t+2x\partial_x+ (\rho+1)y\partial_y$\\{\smallskip}

11&
$e^t$&$\partial_x,\quad\partial_y,\quad  x\partial_x+\partial_u,\quad2\partial_t+y\partial_y$\\{\smallskip}

12&
1&$\partial_x,\quad\partial_y,\quad x\partial_x+\partial_u,\quad\partial_t,\quad2t\partial_t+2x\partial_x+y\partial_y$\\
\hline\noalign{\smallskip}
\multicolumn{3}{c}{$k=\ln u$}\\
\hline\noalign{\smallskip}
13&
$\forall$&$\partial_x,\quad\partial_y,\quad t\partial_x+u\partial_u$\\{\smallskip}

14&
$t^\rho$&$\partial_x,\quad\partial_y,\quad t\partial_x+u\partial_u,\quad2 t\partial_t+2x\partial_x+ (\rho+1)y\partial_y$\\{\smallskip}

15&
$e^t$&$\partial_x,\quad\partial_y,\quad t\partial_x+u\partial_u,\quad2\partial_t+y\partial_y$\\{\smallskip}

16&
1&$\partial_x,\quad\partial_y,\quad t\partial_x+u\partial_u,\quad\partial_t,\quad2t\partial_t+2x\partial_x+y\partial_y$\\
\noalign{\smallskip}\hline\noalign{\smallskip}
\end{tabular}
\mbox{Here $n$ and $\rho$ are arbitrary nonzero constants, and $n\neq1$.}
\end{table}

If $k=u^n$, $n\neq0,1,$  then splitting equations~(\ref{deteq2}) and~(\ref{deteq3}) with respect to different powers of~$u$ leads to the system
$\xi_t=\psi=\varphi_x=0,$ $\varphi_t=f\varphi_{yy}$, $n\varphi+\tau_t-\xi_x=0.$ These equations together with~(\ref{deteq1}) imply $\tau=c_1t+c_2,$ $\xi=(c_1+n c_6)x+c_3$, $\eta=c_4y+c_5,$ $\varphi=c_6$, where $c_i,$ $i=1,\dots,6$, are arbitrary constants. The classifying equation for $f$ takes the form~(\ref{classeq1}). Therefore, the cases of Lie symmetry extensions are given by the same forms of $f$ as in previous case, namely, arbitrary, power, exponential and constant. See Cases 5--8 of Table~1. The dimensions of the respective Lie symmetry algebras increase by one in comparing with the case of arbitrary $k.$ The highest dimension is five, not six as it was stated in the paper by Kumar et al~\cite{Kumar2014}.

The consideration of the cases $k=e^u$ and $k=\ln u$ is rather similar to the case of $k=u^n$ with $n\neq0,1,$ therefore, we omit the details of calculations. The classification results are presented in Cases 9--16 of Table~1.

Consider the case of linear $k$, then up to the equivalence we can assume  $k=u$. We substitute $k=u$ to equations~(\ref{deteq2}) and~(\ref{deteq3}) and further split them with respect to different powers of~$u$. This leads to the system
$\psi   =\xi_t,$ $ \tau_t-\xi_x+\varphi=0,$ $\varphi_x=0,$
$\psi_x  +\varphi_t-f\varphi_{yy}=0,$ and $\psi_t-f\psi_{yy}=0.$
We differentiate the first and the second equation of this system with respect to the variable~$y$ and get the additional conditions $\varphi_y=\psi_y=0.$
Then  also $\psi_t=\psi_{xx}=\varphi_{tt}=0 $ and the second equation of~(\ref{deteq1})  gives $\eta^1_t=\eta^0_t=0.$
The general form of the infinitesimal operator $Q$ is
$
Q=(c_2t^2+c_1t+c_0)\partial_t+((c_2t+c_4)x+c_3t+c_5)\partial_x+ (c_6
y+c_7)\partial_y+((c_4-c_1-c_2t)u+c_2x+c_3)\partial_u,
$
where $c_i,$ $i=0,\dots,7,$ are arbitrary constants.
The classifying equation for $f$ is
\begin{eqnarray}\label{eq_classifying}
(c_2t^2+c_1t+c_0)f _t=(2c_6-c_1-2c_2t)f.
\end{eqnarray}

If this is not an equation on $f$ but an identity, then $c_0=c_1=c_2=c_6=0.$ Therefore, the constants $c_3,$ $c_4,$ $c_5,$ $c_7$ appearing in the infinitesimal generator $Q$ are arbitrary and the maximal Lie invariance algebra of the equations~(\ref{eq_2dDiff5}) with arbitrary $f$ is the four-dimensional algebra $\langle \partial_x,\ \partial_y,\   x\partial_x+u\partial_u,\  t\partial_x+\partial_u\rangle$ (Case 1 of Table~2).

\begin{table}[t!]\centering
\caption{The group classification of class~(\ref{eq_2dDiff5}) up to the $G^\sim_1$-equivalence.}
\begin{tabular}{ccl}
\hline\noalign{\smallskip}
no.&$f (t)$&\hfil Basis of $A^{\max}$ \\
\noalign{\smallskip}\svhline\noalign{\smallskip}
1&
$\forall$&$\partial_x,\quad\partial_y,\quad  x\partial_x+u\partial_u,\quad t\partial_x+\partial_u$\\{\smallskip}
2&
$\displaystyle \frac{e^{\sigma\arctan t}}{t^2+1}$&$\partial_x,\quad\partial_y,\quad  x\partial_x+u\partial_u,\quad t\partial_x+\partial_u,\quad(t^2+1)\partial_t+tx\partial_x+\frac12\sigma y\partial_y+(x-tu)\partial_u$\\{\smallskip}
3&
$t^\rho$&$\partial_x,\quad\partial_y,\quad  x\partial_x+u\partial_u,\quad t\partial_x+\partial_u,\quad2t\partial_t+(\rho+1)y\partial_y-2u\partial_u$\\{\smallskip}
4&
$e^t$&$\partial_x,\quad\partial_y,\quad  x\partial_x+u\partial_u,\quad t\partial_x+\partial_u,\quad2\partial_t+y\partial_y$\\{\smallskip}
5&
$1$&$\partial_x,\quad\partial_y,\quad  x\partial_x+u\partial_u,\quad t\partial_x+\partial_u,\quad\partial_t,\quad 2t\partial_t+y\partial_y-2u\partial_u$\\
\noalign{\smallskip}\hline\noalign{\smallskip}
\end{tabular}
\mbox{Here $\rho$ and $\sigma$ are arbitrary constants with $\rho\neq0,-2$. Moreover $\rho\leq-1\bmod G_1^\sim$.}
\end{table}

The further group classification of equations~(\ref{eq_2dDiff4}) with $k=u$, i.e. equations~(\ref{eq_2dDiff5}), is equivalent to the integration of the following equation on $f$
\begin{eqnarray}\label{eq_classifyingA}(at^2+bt+c)f_t=(d-2at)f,\end{eqnarray}
where $a,$ $b,$ $c$ and  $d$ are arbitrary constants with $(a,b,c)\neq(0,0,0).$
Up to $G^{\sim}_1$-equivalence
the parameter quadruple~$(a,b,c,d)$ can be assumed to belong to the set
$
\{(1,0,1,\sigma),\ (0,1,0,\rho),\ (0,0,1,1),\ (0,0,1,0)\},
$
where $\sigma$, $\rho$ are nonzero constants, $\rho\leq-1$.
The proof is similar to ones presented in Vaneeva et al.~\cite{VPS2012,VSL2015}. It is based on the fact that transformations from the equivalence group $G^\sim_1$ can be extended to the coefficients $a$, $b$, $c$ and $d$ as follows
\begin{eqnarray*}
\begin{array}{l}
\tilde a=\mu(a\delta^2-b\gamma\delta+c\gamma^2),\quad
\tilde b=\mu(-2a\beta\delta+b(\alpha\delta+\beta\gamma)-2c\alpha\gamma),
\\[1ex]
\tilde c=\mu(a\beta^2-b\alpha\beta+c\alpha^2),\quad
\tilde d=\mu (d\Delta+2a\beta\delta-2b\beta\gamma+2c\alpha\gamma),
\end{array}
\end{eqnarray*}
where $\Delta=\alpha\delta-\beta\gamma\neq0$ and $\mu$ is an arbitrary nonzero constant.

Integration of the equation~(\ref{eq_classifyingA}) for four inequivalent cases of the quadruple $(a,b,c,d)$  gives respectively $f=\frac{e^{\sigma\arctan t}}{t^2+1},$ $f=t^\rho,$ $\rho\neq0$, $f=e^t$ and $f=1.$ We further substitute the obtained inequivalent values of $f$ into equation~(\ref{eq_classifying}) and find the corresponding values of constants $c_i$ and, therefore, the general forms of the infinitesimal generators. The results of the group classification of class~(\ref{eq_2dDiff5}) are presented in Table~2.

The classification lists presented in Tables~1 and 2 give the exhaustive group classification of the class of variable coefficient nonlinear Kolmogorov equations~(\ref{eq_2dDiff2}) with nonlinear $k$ and of the class of equations~(\ref{eq_2dDiff3}) up to the $\hat G^{\sim}$- and $\hat G^{\sim}_1$-equivalences, respectively.

\section{Discussion on the choice of the optimal gauging}

Appropriate choice of gauging of the arbitrary elements is a crucial step in solving group classification problems.
The gauging $f=1$ could seem more convenient if one look for the determining equations for finding Lie symmetries.
For  class~(\ref{eq_2dDiff7})
 they have the form
\begin{eqnarray*}
&2\eta_y=\tau_t,\quad
\eta_{yy}-\eta_t=2\varphi_y,\quad
(\varphi u+\psi)g  k_u+[\tau g_t+(\tau_t-\xi_x)g ]k=\xi_t,\\&
(\varphi_x u+\psi_x)g  k +(\varphi_t-\varphi_{yy})u+\psi_t-\psi_{yy}=0.
\end{eqnarray*}
For the case $k\neq u$ the difference in classification is not so crucial (cf. Table~1 with Table~3). Though one can see that
for $k=\ln u$ the operator $t\partial_x+u\partial_u$ appearing in Cases 13--16 of Table~1 transforms to various forms in the respective cases of Table~3. For the case $k=u$ the difficulty of group classification of the class~(\ref{eq_2dDiff2}) with $f=1$ increases essentially in comparison with the gauging \mbox{$g=1$}. Solving the determining equations results in the following form of the infinitesimal generator
\begin{eqnarray*}\textstyle Q=(c_1t+c_0)\partial_t+[(c_2x+c_3)\int\! g (t)\,{\rm d}t+c_4x+c_5]\partial_x+{}\\\qquad\ \textstyle(\frac12c_1y+c_6)\partial_y+[(c_7-c_2\int\! g (t)\,{\rm d}t)u+c_2x+c_3]\partial_u,\end{eqnarray*}
where $c_i$, $i=0,\dots,7,$ are arbitrary constants. The classifying equation for $g$ is the integro-differential equation $ (c_1t+c_0) g _t+\left(c_1-c_4+c_7-2c_2\int\! g (t)\,{\rm d}t\right) g =0$ (cf. with the classifying equation~(\ref{eq_classifying}) for $f$ that is much simpler). The results of group classification for class~(\ref{eq_2dDiff6})
are presented in Table~4. Comparing Tables~2 and~4 one can conclude that forms of the basis operators of the maximal Lie invariance algebras are more cumbersome in Table~4.
\begin{table}[t!]\centering
\caption{The group classification of class~(\ref{eq_2dDiff7})  up to the $\hat G^\sim_2$-equivalence.}
\begin{tabular}{ccl}
\hline\noalign{\smallskip}
no.&$g (t)$&\hfil Basis of $A^{\max}$ \\
\noalign{\smallskip}\svhline\noalign{\smallskip}
\multicolumn{3}{c}{Arbitrary $k$}\\
\hline\noalign{\smallskip}
1&
$\forall$&$\partial_x,\quad\partial_y$\\{\smallskip}

2&
$t^\rho$&$\partial_x,\quad\partial_y,\quad2t\partial_t+2(\rho+1)x\partial_x+y\partial_y$\\{\smallskip}

3&
$e^t$&$\partial_x,\quad\partial_y,\quad\partial_t+x\partial_x$\\{\smallskip}

4&
$1$&$\partial_x,\quad\partial_y,\quad\partial_t,\quad2t\partial_t+2x\partial_x+y\partial_y$\\
\hline\noalign{\smallskip}
\multicolumn{3}{c}{$k=u^{n}$, ${n}\neq0,1$}\\
\hline\noalign{\smallskip}
5&
$\forall$&$\partial_x,\quad\partial_y,\quad{n} x\partial_x+u\partial_u$\\{\smallskip}

6&
$t^\rho$&$\partial_x,\quad\partial_y,\quad{n} x\partial_x+u\partial_u,\quad2 t\partial_t+2(\rho+1)x\partial_x+ y\partial_y$\\{\smallskip}

7&
$e^t$&$\partial_x,\quad\partial_y,\quad{n} x\partial_x+u\partial_u,\quad\partial_t+x\partial_x$\\{\smallskip}

8&
1&$\partial_x,\quad\partial_y,\quad{n} x\partial_x+u\partial_u,\quad\partial_t,\quad2t\partial_t+2x\partial_x+y\partial_y$\\
\hline\noalign{\smallskip}
\multicolumn{3}{c}{$k=e^u$}\\
\hline\noalign{\smallskip}
9&
$\forall$&$\partial_x,\quad\partial_y,\quad  x\partial_x+\partial_u$\\{\smallskip}

10&
$t^\rho$&$\partial_x,\quad\partial_y,\quad  x\partial_x+\partial_u,\quad2 t\partial_t+2(\rho+1)x\partial_x+ y\partial_y$\\{\smallskip}

11&
$e^t$&$\partial_x,\quad\partial_y,\quad  x\partial_x+\partial_u,\quad\partial_t+x\partial_x$\\{\smallskip}

12&
1&$\partial_x,\quad\partial_y,\quad x\partial_x+\partial_u,\quad\partial_t,\quad2t\partial_t+2x\partial_x+y\partial_y$\\
\hline\noalign{\smallskip}
\multicolumn{3}{c}{$k=\ln u$}\\
\hline\noalign{\smallskip}
13&
$\forall$&$\partial_x,\quad\partial_y,\quad \int\! g (t)\,{\rm d}t\,\partial_x+u\partial_u$\\{\smallskip}

$14_a$&
$t^\rho,\,\rho\neq-1$&$\partial_x,\quad\partial_y,\quad t^{\rho+1}\partial_x+(\rho+1)u\partial_u,\quad2 t\partial_t+2(\rho+1)x\partial_x+ y\partial_y$\\{\smallskip}

$14_b$&
$t^{-1}$&$\partial_x,\quad\partial_y,\quad\ln t\,\partial_x+u\partial_u,\quad2 t\partial_t+ y\partial_y$\\{\smallskip}
15&
$e^t$&$\partial_x,\quad\partial_y,\quad e^t\partial_x+u\partial_u,\quad\partial_t+x\partial_x$\\{\smallskip}

16&
1&$\partial_x,\quad\partial_y,\quad t\partial_x+u\partial_u,\quad2t\partial_t+2x\partial_x+y\partial_y,\quad\partial_t$\\
\noalign{\smallskip}\hline\noalign{\smallskip}
\end{tabular}
\mbox{Here $n$ and $\rho$ are arbitrary nonzero constants.}
\end{table}
\begin{table}[ht!]\centering
\caption{The group classification of class~(\ref{eq_2dDiff6}) up to the $\hat G^\sim_3$-equivalence.}
\begin{tabular}{ccl}
\hline\noalign{\smallskip}
no.&$g (t)$&\hfil Basis of $A^{\max}$ \\
\noalign{\smallskip}\svhline\noalign{\smallskip}
1&
$\forall$&$\partial_x,\quad\partial_y,\quad  x\partial_x+u\partial_u,\quad \int\!  g (t)\,{\rm d}t\;\partial_x+\partial_u$\\{\smallskip}

2&
$\displaystyle\frac1{t\cos^2(\nu\ln t)}$&$\partial_x,\quad\partial_y,\quad x\partial_x+u\partial_u,\quad \tan(\nu\ln t)\partial_x+\nu\partial_u,$\\ &&$ \ t\partial_t+\nu x\tan(\nu\ln t)\partial_x+ \frac12y\partial_y+\nu(\nu x-\tan(\nu\ln t)u)\partial_u$\\{\smallskip}

3&
$\displaystyle\frac1{\cos^2t}$&$\partial_x,\quad\partial_y,\quad x\partial_x+u\partial_u,\quad \tan t\partial_x+\partial_u,\quad \partial_t+x\tan t\partial_x+( x-u\tan t)\partial_u$\\{\smallskip}

$4_a$&
$t^\rho$&$\partial_x,\quad\partial_y,\quad x\partial_x+u\partial_u,\quad t^{\rho+1}\partial_x+(\rho+1)\partial_u,\quad2 t\partial_t+2(\rho+1)x\partial_x+ y\partial_y$\\{\smallskip}

$4_b$&
$t^{-1}$&$\partial_x,\quad\partial_y,\quad x\partial_x+u\partial_u,\quad \ln t\,\partial_x+\partial_u,\quad2 t\partial_t+ y\partial_y$\\{\smallskip}

5&
$e^t$&$\partial_x,\quad\partial_y,\quad  x\partial_x+u\partial_u,\quad e^t\partial_x+\partial_u,\quad\partial_t+x\partial_x$\\{\smallskip}

6&
1&$\partial_x,\quad\partial_y,\quad x\partial_x+u\partial_u,\quad t\partial_x+\partial_u,\quad\partial_t,\quad2t\partial_t+2x\partial_x+y\partial_y$\\
\noalign{\smallskip}\hline\noalign{\smallskip}
\end{tabular}
\mbox{Here $\rho$ and $\nu$ are arbitrary constants with $\nu\neq0,$ $\rho\neq-2,-1,0.$ Moreover  $\rho<-1\bmod \hat G_3^\sim$.}
\end{table}

 The links between equations of the form~(\ref{eq_2dDiff6}) are also more tricky than those between equations from class~(\ref{eq_2dDiff5}).
For example, the equation
$$
u_t=u_{yy}-\frac1{t\cosh^2(\nu\ln t)}uu_x,
$$
where the variable coefficient  can be rewritten as $\frac4{t(t^\nu+t^{-\nu})^2}$, admits the five-dimensional maximal Lie invariance algebra with the basis operators
 $
\partial_x,$  $\partial_y,$ $\tanh(\nu\ln t)\partial_x+\nu\partial_u,$  $x\partial_x+u\partial_u,$  and $t\partial_t-\nu x\tanh(\nu\ln t)\partial_x+ \frac12y\partial_y-\nu(\nu x-\tanh(\nu\ln t)u)\partial_u.$
 The equivalence of this equation and the equation $$\tilde u_{\tilde t}=\tilde u_{\tilde y\tilde y}-\tilde t^{2\nu-1} \tilde u\tilde u_{\tilde x}$$ from the same class does not seem obvious. Nevertheless, there exists the transformation from the equivalence group $\hat G^\sim_3$,
 \[
\tilde t=t,\quad \tilde x=\frac1{4}{x(t^{2\nu}+1)},\quad \tilde y=y,\quad \tilde u=\frac u{t^{2\nu}+1}+\frac\nu2x,
\]
 that establishes a~link between these equations.
This shows that the distinguishing inequivalent cases of Lie symmetry extensions for class~(\ref{eq_2dDiff6}) is also a more difficult task than for class~(\ref{eq_2dDiff5}).

Therefore, the gauging $g=1$ is without a  doubt the right choice to perform a~group classification for the class~(\ref{eq_2dDiff2}) and especially for its subclass~(\ref{eq_2dDiff3}).
So, is there a regular way that can help one to choose a preferable gauging among several possible ones? Equivalence groups appear to be indicators showing the right choice of gauging. Indeed,  the comparison of the equivalence groups presented in Theorems 3 and 4 with those given in Theorems 5 and 6 shows that the equivalence groups of class~(\ref{eq_2dDiff4}) and its subclass~(\ref{eq_2dDiff5}) are usual  whereas the equivalence  groups  of  class~(\ref{eq_2dDiff7})
 and its subclass~(\ref{eq_2dDiff6}) remain to be generalized extended as the equivalence group of the initial class. Transformations from the generalized extended groups become point only after fixing arbitrary elements and integrals of $g$ then naturally appear in the forms of Lie symmetry generators and even in the classifying equation. This of course makes the calculations more difficult.
Therefore, the widest possible equivalence group should be necessarily found even before applying the Lie invariance criterion to equations under study in order to choose the optimal gauging and to optimize the entire process of group classification.

%\vspace{-3mm}
\section{Conclusion} The complete group classification of class~(\ref{eq_2dDiff}) has been performed  using the gauging of arbitrary elements by  the equivalence transformations.
We have presented classification lists for an equivalent form of this class, namely, for class~(\ref{eq_2dDiff2}). The
corresponding values of  $K$ for the values of
 $k$ from Tables 1 and 3  are the following:
$k=u^n$, $n\neq0,-1$, $\leftrightarrow$ $K=u^{n+1}$; $k=u^{-1}$  $\leftrightarrow$ $K=\ln u$; $k=e^u$  $\leftrightarrow$ $K=e^u$; $k=\ln u$  $\leftrightarrow$ $K=u\ln u$.

Application of the widest possible (generalized extended) equivalence groups allowed us to write down classification lists in the explicit and concise form.
We have also shown that the equivalence group is that indicator that helps one to choose the optimal gauging among several possible ones.

The derived Lie symmetries can be now used to reduce the nonlinear Kolmogorov  equations~(\ref{eq_2dDiff}) to ordinary differential equations and, therefore, for finding exact solutions. The reductions can be achieved using two-dimensional subalgebras of the corresponding maximal Lie invariance algebras.

\begin{acknowledgement}
 O.\,V. would like to thank  the Organizing Committee of LT-11  and especially Prof. Vladimir Dobrev for the hospitality and support.
O.\,V. and Yu.\,K. also acknowledge the support provided by the University of Cyprus.
The authors express their gratitude to Roman Popovych for useful discussions and to the anonymous referee for valuable comments.
\end{acknowledgement}

\end{document}